\documentclass{TGClass}
\usepackage[square]{TGsty}          

\usepackage{amsmath, amssymb}

\usepackage{pstricks}

\makeindex

\newcommand{\GasenzerCmdeq}[1]{(\ref{GasenzerRef:eq:#1})}
\newcommand{\GasenzerCmdEq}[1]{Eq.~(\ref{GasenzerRef:eq:#1})}

\newcommand{\GasenzerCmdFig}[1]{Fig.~\ref{GasenzerRef:fig:#1}}

\begin{document}

\chapter[Functional-integral approaches to quantum many-body dynamics]{Non-equilibrium Quantum Many-Body Dynamics: Functional Integral Approaches \label{ch1}}

\author[C. Bodet, M. Kronenwett, B. Nowak, D. Sexty, and T. Gasenzer]
{C\'edric Bodet, Matthias Kronenwett, Boris Nowak, D\'{e}nes Sexty, and Thomas Gasenzer\footnote{t.gasenzer@uni-heidelberg.de}}

\address{Institut f\"ur Theoretische Physik, Universit\"at Heidelberg,\\
Philosophenweg 16, 69120 Heidelberg, Germany, and \\
	    ExtreMe Matter Institute EMMI,
             GSI Helmholtzzentrum f\"ur Schwerionenforschung GmbH, 
             Planckstra\ss e~1, 
             D--64291~Darmstadt
}

\begin{abstract}
We discuss functional-integral approaches to far-from-equilibrium quantum many-body dynamics.
Specific techniques considered include the two-particle-irreducible effective action and the real-time flow-equation approach.
Different applications, including equilibration after a sudden parameter change and non-equilibrium critical phenomena, illustrate the potential of these methods.
\end{abstract}

\body

\section{Introduction}

Time evolution of many-body systems in which effects of quantum physics play an important role belong to the least understood physical phenomena. 
Quantum effects prevent a fully deterministic description of time evolution and they render the problem mathematically difficult. 
Two main aspects constitute the complexity connected with non-equilibrium phenomena and their theoretical description: 
long-time evolution and strong correlations. 
These aspects can not be considered independently from each other.
Due to the enormous progress in the field of ultracold atomic quantum gases in recent years, quantum many-body dynamics can now be very precisely controlled and probed and has become a timely topic for laboratory studies. 
On the theory side, the exponential growth of available computing power has brought quantum dynamics of many-body systems into the reach of large-scale calculations while sophisticated advanced analytical approaches open the way to a deeper understanding of quantum many-body evolution.

In situations close to equilibrium, one can treat non-equilibrium dynamics as a perturbation of the equilibrium state. 
In different ways, most presently available field-theoretical methods rely to a certain extent on a perturbation expansion and either require a weak coupling or break down at large times of the evolution. 
This feature is inherent in principle, and it poses a serious problem when applying present results to quantum many-body dynamics far from equilibrium or to strongly coupled systems. 
The escape from this restriction is generally seen in refined approximations which allow to take complex correlations into account over a sufficiently long period of evolution before they cease haveing any further influence.
Finding suitable approximations beyond leading-order perturbation theory can become technically involved. 
In this respect, functional-integral approaches including the two-particle irreducible (2PI) effective-action and the real-time functional flow equation methods for quantum field dynamics represent successful ways of optimizing outcome versus effort.
A central feature of these approaches is that the resulting dynamical equations conserve crucial quantities like energy, irrespective of the chosen approximation. 
Most importantly, the methods allow approximations beyond an expansion in powers of the interaction strength. 
This review summarizes briefly the main aspects of the formalism and presents three examples for applications in which we compare with other methods and give specific predictions amenable to experimental investigation.

\section{Non-equilibrium quantum field theory}
\label{GasenzerRef:sec:NEqFT}
Assuming basic knowledge about Feynman path integrals, we sketch the functional-integral formulation of real-time quantum field theory. 
To derive conserving many-body dynamic equations beyond mean-field order, we introduce the two-particle irreducible effective action as well as the real-time flow-equation approach.

\subsection{Schwinger-Keldysh closed time path}
\label{GasenzerRef:sec:CTP}
We will consider initial-value problems, assuming that the many-body state is given  by a density matrix $\rho(t_0)$ at some initial time $t=t_0$.
The Schr\"odinger- and Heisenberg-picture evolutions of expectation values are related to each other by the time evolution operator  $U(t,t')={\cal T}\exp\{-i\int_{t'}^t \mathrm{d}t''\,H(t'')\}$ determined by the Hamiltonian $H$ (we use $\hbar=1$),
\begin{equation}
\label{GasenzerRef:eq:Ot}
   \langle{\cal O}(t)\rangle=\mathrm{Tr}\left[\rho(t){\cal O}(t)\right]
  = \mathrm{Tr}\left[\rho(t_0)U^\dagger(t,t_0){\cal O}(t)U(t,t_0)\right].
\end{equation}
We use four-vector notation $x=(x_{0},\mathbf{x})$ with time $t=x_{0}$ and space coordinate $\mathbf{x}$.
The operator ${\cal O}$ is usually a product of field operators evaluated at different times such that, e.g., taking $x_{0}>y_{0}$,
\begin{equation}
\label{GasenzerRef:eq:Gtwotime}
  \langle\Phi(x)\Phi(y)\rangle
  =\mathrm{Tr}\left[\rho(t_0)\,U^\dagger(x_0,t_0)\Phi(\mathbf{x})U(x_0,y_0)\Phi(\mathbf{y})U(y_0,t_0)\right].
\end{equation}
%
%
\begin{figure}[tb]
\begin{center}
\resizebox{0.6\columnwidth}{!}{
\includegraphics{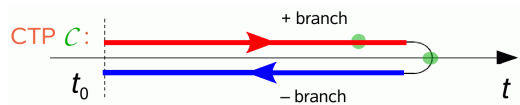}
}
\end{center}
\vspace*{-3ex}
\caption{
Schwinger-Keldysh closed time path ${\cal C}$. 
The green dots indicate the times $x_0$ and $y_0$ for an example two-point function $\langle\Phi(x)\Phi(y)\rangle$, see text.
The branches are drawn above and below the time axis to make them separately visible.}
\label{GasenzerRef:fig:CTP}
\end{figure}
$\!\!$The product of different time evolution operators and field operators evaluates along a closed time path (CTP), \GasenzerCmdFig{CTP}.
If one chooses all operator times to lie on the $+$ branch, the path integral gives the expectation value of the time-ordered product.
In the path-integral approach, all evolution operators $U$ are replaced by path integrals.
Hence, the path integral consists of a product of integrals, one for each point along the CTP.

\subsection{Generating functional for correlation functions}
\label{GasenzerRef:sec:GenFunc}
Within the Schwinger-Keldysh approach, the fundamental entity of non-equilibrium dynamics is the trace over the initial density operator multiplied by the forward and backward time evolution operators.
Rewriting the latter as path integrals, the generating functional for correlation functions reads 
\begin{equation}
\label{GasenzerRef:eq:ZJ}
  Z_{\rho_{0}}[J] 
  = \mathrm{Tr}\left[\rho(t_{0})\mathcal{T}_{\mathcal{C}}e^{i\int_{\mathcal{C}} J\Phi}\right] 
  = \int{\cal D}\varphi\, \rho[\varphi^+_0,\varphi^-_0] e^{i(S_{\mathcal{C}}[\varphi]+\int_{\mathcal{C}} J\varphi)},
\end{equation}
with $\rho[\varphi^+_0,\varphi^-_0]=\langle\varphi_0^{+}|\rho(t_{0})|\varphi_0^{-}\rangle$.
The external source field $J(x)$ turns the path integral into a generating functional for correlation functions, similarly as in the (grand) canonical partition function in equilibrium physics, and $\mathcal{T}_{\mathcal{C}}$ denotes operator ordering along the CTP $\mathcal{C}$. 
Furthermore, $\int_{\mathcal{C}} J\varphi=\int_{\cal C} \mathrm{d}^{d+1}x J(x)\varphi(x)=\int \mathrm{d}^dx[\int_{t_0}^{t_\mathrm{max}}\mathrm{d}x_0 J^{+}(x)\varphi^{+}(x)+\int_{t_\mathrm{max}}^{t_0}\mathrm{d}x_0 J^{-}(x)\varphi^{-}(x)]$, and ${\cal D}\varphi=\prod_{x}[\mathrm{d}\varphi^{-}(x)][\mathrm{d}\varphi^{+}(x)]$, 
with $[\mathrm{d}\varphi^\pm]=\prod_{\mathbf{x}}\mathrm{d}\varphi^\pm(t_0,\mathbf{x})$.
$|\varphi_{0}^{\pm}\rangle$ are eigen states of $\Phi(t_{0},\mathbf{x})$, and
$S_{\mathcal{C}}$ is a sum of two actions such that the overall time integral runs along the CTP.
The generating functional allows, e.g., the field expectation value $\phi(x)=\langle\Phi(x)\rangle$ to be written as
\begin{equation}
\label{GasenzerRef:eq:phifromZ}
  \phi(x) = -i \left.\frac{\delta \ln Z_{\rho_0}[J]}{\delta J(x)}\right|_{J=0}
  = Z_{\rho_{0}}^{-1}\int{\cal D}\varphi\, \varphi(x)\,\rho[\varphi^+_0,\varphi^-_0]\,e^{iS_{\mathcal{C}}[\varphi]}.
\end{equation}
Due to causality, the CTP extends only to the maximum time to be evaluated in a particular $n$-point function.
At later times, the sources can be set to zero such that the time evolution operators on the corresponding $+$ and $-$ branches cancel by unitarity.

\subsection{The two-particle irreducible (2PI) effective action}
\label{GasenzerRef:sec:2PIEA}
The 2PI effective action is obtained by a double Legendre transform of $-i\ln Z_{\rho_{0}}$ with respect to $J$ and to a further two-point source $R_{ab}(x,y)$,
\begin{equation}
  \Gamma[\phi,G]
  =\Gamma^R[\phi]-\frac{1}{2}\int_{xy}R_{ab}(x,y)[\phi_b(y)\phi_a(x)+G_{ba}(y,x)]
\label{GasenzerRef:eq:LTrafoGammaR}
\end{equation}
where $\int_x \equiv \int_{\mathcal{C}} \mathrm{d} x_0 \int \mathrm{d}^d x$, and $\Gamma^R[\phi] = W^R[J]-\int J_a\phi_a$ is the first Legendre transform of the Schwinger functional $W^R=-i\ln Z_{\rho_{0}}^R$ for the modified classical action $S_{\mathcal{C}}^R[\varphi]=S_{\mathcal{C}}[\varphi]+\int_{\mathcal{C}}\varphi R\varphi/2$.
The classical source fields $J$ and $R$ act as Lagrange multipliers for the extremization of the 2PI effective action with respect to $\phi$ and the two-point Green function $G_{ab}(x,y)=\langle{\cal T_C}\Phi_a(x)\Phi_b(y)\rangle-\phi_a(x)\phi_b(y)$, under the constraints \GasenzerCmdeq{phifromZ}  and $\delta^2 W^R/\delta J_a(x)\delta J_b(y)|_{J=0}=G_{ab}(x,y)$.
In turn, the action functional \GasenzerCmdeq{LTrafoGammaR} allows to determine the correlation functions $\phi$ and $G$
from the Hamiltonian variational conditions
\begin{equation}
\label{GasenzerRef:eq:StatCondsphiG}
    \frac{\delta\Gamma[\phi,G]}{\delta\phi_a(x)}
  = -J_a(x)-\int_{y,\mathcal{C}} R_{ab}(x,y)\phi_b(y),~~~
  \frac{\delta\Gamma[\phi,G]}{\delta G_{ab}(x,y)}
  = -\frac{1}{2}R_{ab}(x,y).
\end{equation}

Presuming a non-vanishing $\phi$, a saddle-point expansion can be used to write the 2PI effective action
in terms of a one-loop part and a ``rest'' $\Gamma_2$,
\begin{equation}
\label{GasenzerRef:eq:Gamma2PI1loop}
  \Gamma[\phi,G]
  = S_{\mathcal{C}}[\phi]+\frac{i}{2}\mathrm{Tr}\left(\ln G^{-1}+G_0^{-1}G\right)+\Gamma_2[\phi,G]+\mathrm{const.},
\end{equation}
with free inverse propagator $iG^{-1}_{0,ab}(x,y)={\delta^2 S_{\mathcal{C}}[\phi]}/{\delta\phi_a(x)\delta\phi_b(y)}$.
For $R=0$, $\Gamma_2$ yields the self-energy
\begin{equation}
\label{GasenzerRef:eq:SigmafromGamma2}
  \Sigma_{ab}(x,y;\phi,G)=2i\frac{\delta\Gamma_2[\phi,G]}{\delta G_{ab}(x,y)}
\end{equation}
that contains all information about scattering in the dynamics. 
The dynamical evolution of $G$ is governed by the Dyson equation obtained from the second equation in \GasenzerCmdeq{StatCondsphiG}:
\begin{equation}
\label{GasenzerRef:eq:DefSigma}
  G^{-1}(x,y)
  = G_0^{-1}(x,y)-\Sigma(x,y).
\end{equation}
Upon multiplication by $G$ this becomes a time evolution equation for the two-point Green function $G$,
\begin{equation}
\label{GasenzerRef:eq:EOMG}
  \int_zG^{-1}_{0,ac}(x,z)G_{cb}(z,y) 
  = \delta_{ab}\delta_{\cal C}(x-y)+\int_z\Sigma_{ac}(x,z)G_{cb}(z,y),
\end{equation}
with $\delta_{\cal C}(x-y)=\delta_{\cal C}(x_0-y_0)\delta(\mathbf{x}-\mathbf{y})$.
A corresponding equation can be obtained for $\phi_a$.
%
\begin{figure}[tb]
\begin{center}
\begin{minipage}{0.35\columnwidth}
\resizebox{1.0\columnwidth}{!}{
\includegraphics{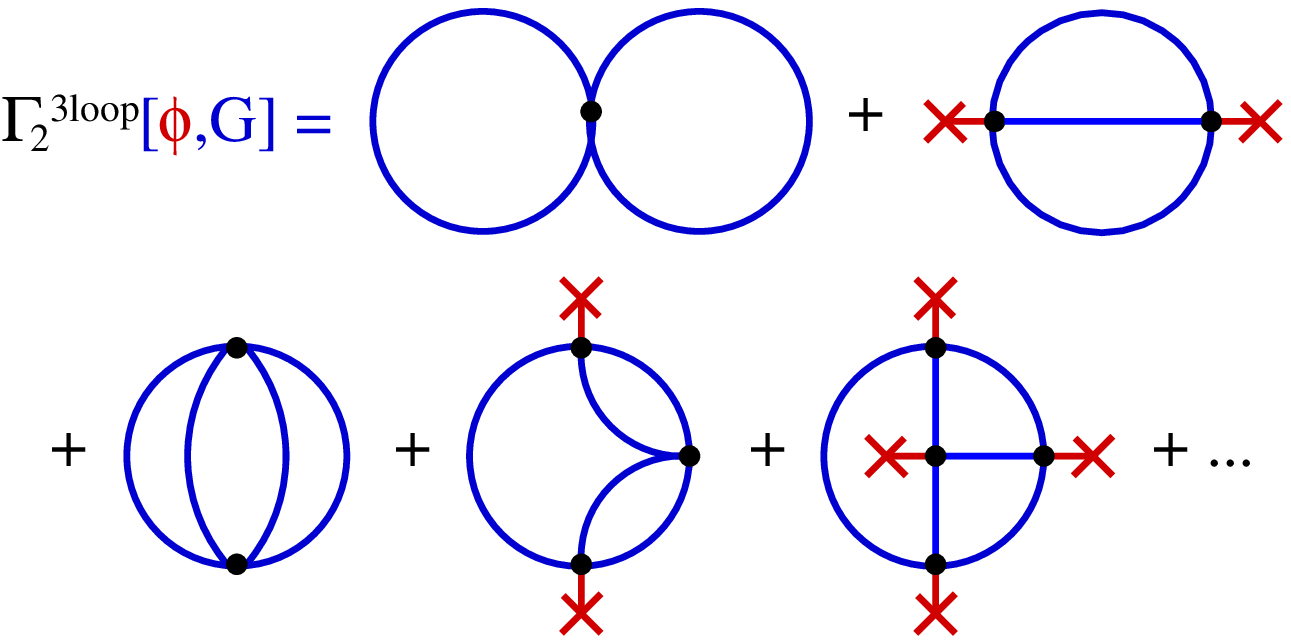}
}
\end{minipage}
\hspace*{0.05\columnwidth}
\begin{minipage}{0.55\columnwidth}
\resizebox{1.0\columnwidth}{!}{
\includegraphics{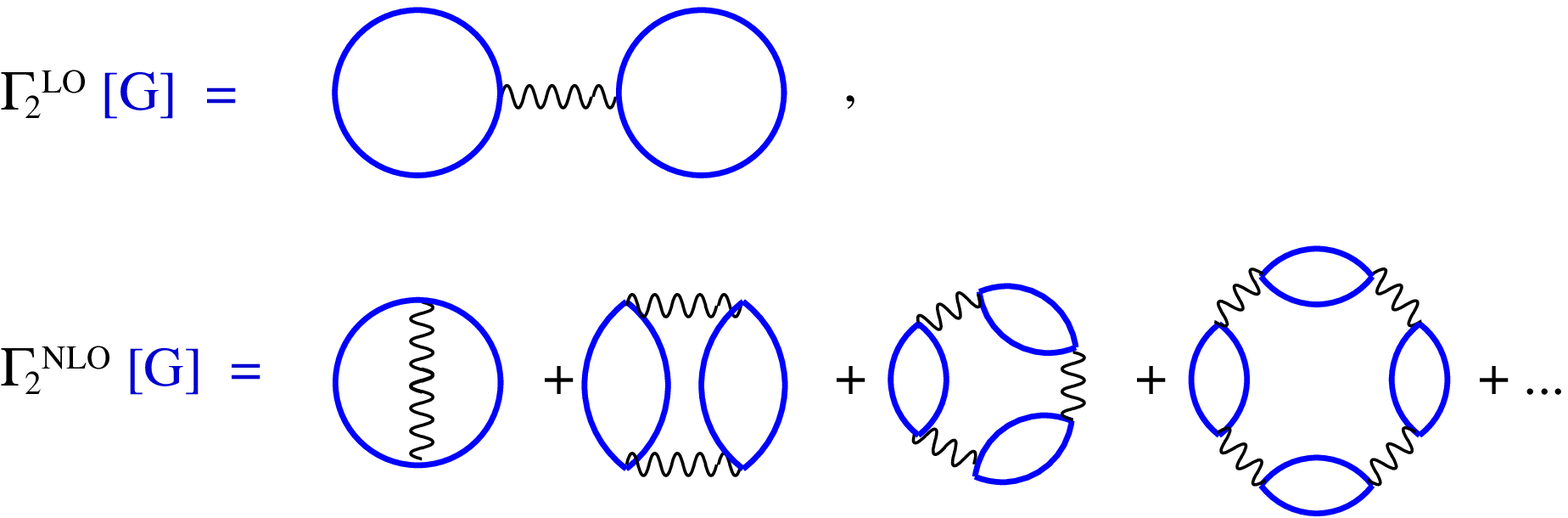}
}
\end{minipage}
\end{center}
\caption{
(Left panel) Diagrammatic representation of the two- and three-loop diagrams contributing to $\Gamma_2[\phi,G]$ in the 2PI effective action, for a system with elastic two-to-two collisions.
The bare interaction vertices are drawn as black dots.
(Right panel) Diagrammatic representation of the  leading order (LO) and next-to-leading order (NLO) contributions in the $1/\cal N$-expansion, to the 2PI part $\Gamma_2[\phi,G]$ of the 2PI effective action, for $\phi=0$.
The thick blue lines represent 2-point functions $G_{ab}(x,y)$ and the wiggly lines a single vertex channel.
At each end of the wiggly lines, it is summed over the field indices of the $G$ lines  ending there, and at the whole vertex it is integrated over time and space.
}
\label{GasenzerRef:fig:Gamma23loop}
\end{figure}
Since the self-energy $\Sigma$ is one-particle irreducible, and since taking the derivative with respect to $G$ corresponds to opening a propagator line, it follows that $\Gamma_2$ must consist of closed 2PI diagrams only, those which do not fall apart on opening two propagator lines, see \GasenzerCmdFig{Gamma23loop}.
This forms the central result that the 2PI effective action is given, besides the terms \GasenzerCmdeq{Gamma2PI1loop}, by a series of all closed 2PI diagrams which can be formed from the full propagator $G$, the bare vertices defined by the classical action, and at most two external field insertions $\phi$.
We conclude that the 2PI effective action approach yields a closed set of dynamic equations for $\phi$ and $G$ provided that the correlations in the initial state $\rho_{0}$ can be fully encoded in these functions.
This is the case for Gaussian initial conditions for which all $n$-point correlation functions with $n\ge3$ can be expressed in terms of $\phi$ and $G$.%
\footnote{%
For non-Gaussian initial states
a straightforward generalization of the approach involving the $n$PI effective action is at hand \protect\cite{GasenzerRef:Berges:2004pu}. }
Such higher-order correlations, however, build up during the evolution and are implicitly accounted for in the dynamic equations for $\phi$ and $G$.

The 2PI effective action has been introduced to solid-state theory (there called $\Phi$-functional) \cite{GasenzerRef:Luttinger1960a,GasenzerRef:Baym1962a} and later to relativistic quantum field theory \cite{GasenzerRef:Cornwall1974a}.
See also Refs.~\cite{GasenzerRef:Vasiliev1998a,GasenzerRef:Kleinert1982a}.
Applications to scalar relativistic as well as gauge theories can be found in Refs.~\cite{GasenzerRef:Berges:2001fi,GasenzerRef:Aarts:2002dj,GasenzerRef:Berges:2002cz,GasenzerRef:Mihaila:2000sr,GasenzerRef:Cooper:2002qd,GasenzerRef:Arrizabalaga:2004iw,GasenzerRef:Berges:2002wr,GasenzerRef:Berges:2004ce,GasenzerRef:Berges:2004pu,GasenzerRef:Berges:2004yj,GasenzerRef:Aarts:2006pa,GasenzerRef:Berges:2008wm}, to non-relativistic systems, in particular to ultracold gases, in Refs.~\cite{GasenzerRef:Rey2004a,GasenzerRef:Baier:2004hm,GasenzerRef:Gasenzer:2005ze,GasenzerRef:Rey2005a,GasenzerRef:Temme2006a,GasenzerRef:Berges:2007ym,GasenzerRef:Branschadel:2008sk,GasenzerRef:Scheppach:2009wu,GasenzerRef:Sexty:2010ra}.

\subsection{Functional flow-equation approach}
\label{GasenzerRef:sec:DynFlowEq}
We briefly sketch an alternative approach which is based on functional renormalization group (RG) techniques \cite{GasenzerRef:Wetterich:1992yh,GasenzerRef:Bagnuls:2000ae,
GasenzerRef:Berges:2000ew,
GasenzerRef:Polonyi:2001se,
GasenzerRef:Salmhofer:2001tr,
GasenzerRef:Delamotte:2003dw,
GasenzerRef:Salmhofer:2006pn,
GasenzerRef:Gies:2006wv,
GasenzerRef:Igarashi:2009tj,
GasenzerRef:Rosten:2010vm,
GasenzerRef:Litim:1998nf,
GasenzerRef:Litim:2006ag,
GasenzerRef:Pawlowski:2005xe,
GasenzerRef:Canet:2003yu,
GasenzerRef:Latorre:2004pk,
GasenzerRef:Kehrein2004a,
GasenzerRef:Mitra2006a,
GasenzerRef:Zanella:2006am,
GasenzerRef:Canet:2006xu,
GasenzerRef:Gezzi2007a,
GasenzerRef:Jakobs2007a,
GasenzerRef:Korb2007a,
GasenzerRef:Matarrese:2007wc,
GasenzerRef:Karrasch2008a,
GasenzerRef:Jakobs2009a,
GasenzerRef:Schoeller2009a}. 
For more details on this approach, see Refs.~\cite{GasenzerRef:Gasenzer:2008zz,GasenzerRef:Gasenzer:2010rq}.
Dynamic equations can be derived which are similar in structure to the equations obtained from the 2PI effective action.
The key idea of the approach is to consider the generating functional for Green functions where all times are smaller than a maximum time $\tau$, implying the CTP to be cut off at $\tau$. 
The corresponding generating functional $Z_\tau$ is then defined in terms of the full generating functional $Z=Z_{\rho_{0}}$ \GasenzerCmdeq{ZJ} as
\begin{eqnarray}
\label{GasenzerRef:eq:defZtau}
  Z_\tau 
  &=& \exp\Big\{-\frac{i}{2} \int_{x y}\!
  \frac{\delta}{\delta J_a(x)}R_{\tau,ab}(x,y) \frac{\delta}{\delta
  J_b(y)}\Big\}Z,
\end{eqnarray}
where the function $R_\tau$ is chosen such that it suppresses the fields, i.e., $\delta/\delta J_a$, for all times $t> \tau$. 
This does not fix $R_\tau$ in a unique way, and one choice is 
$-i R_{\tau,ab}(x,y) =   \infty$ for $x_0=y_0>\tau$, $\mathbf{x}=\mathbf{y}$, $a=b$, and zero otherwise, corresponding to $R_{\tau\to\infty}\equiv R=0$ in the 2PI approach.

The time evolution of correlation functions is now derived from that of the Schwinger functional $W_\tau=-i \ln Z_\tau$. 
It is more convenient, however, to consider the time evolution of the effective action
\begin{equation}
  \label{GasenzerRef:eq:effAction}
  \Gamma_\tau[\phi;R_\tau]
  = W_\tau[J;\rho_D]-\int_{\cal C}J_a\phi_a
      - \frac{1}{2}\int_{\cal C}\phi_a  R_{\tau,ab}\phi_b.
\end{equation}
An exact  functional RG or flow equation for the $\tau$-dependent effective action can be derived in the compact form
\begin{equation}
  \label{GasenzerRef:eq:flowGamma}
  \partial_\tau \Gamma_\tau
  = \frac{i}{2} \int_{{\cal C}}\!
  \left[\frac{1}{\Gamma^{(2)}_\tau+R_\tau}\right]_{ab}
  \partial_\tau R_{\tau,ab}\, ,
\end{equation}
where $\Gamma_\tau^{(n)}=\delta^n \Gamma_\tau/(\delta \phi)^n$.
\GasenzerCmdEq{flowGamma} is analogous to the Wetterich functional flow equation \cite{GasenzerRef:Wetterich:1992yh} used extensively with regulators in momentum and/or frequency space to describe strongly correlated systems near equilibrium \cite{GasenzerRef:Wetterich:1992yh,GasenzerRef:Bagnuls:2000ae,
GasenzerRef:Berges:2000ew,
GasenzerRef:Polonyi:2001se,
GasenzerRef:Salmhofer:2001tr,
GasenzerRef:Delamotte:2003dw,
GasenzerRef:Salmhofer:2006pn,
GasenzerRef:Gies:2006wv,
GasenzerRef:Igarashi:2009tj,
GasenzerRef:Rosten:2010vm,
GasenzerRef:Litim:1998nf,
GasenzerRef:Litim:2006ag,
GasenzerRef:Pawlowski:2005xe,
GasenzerRef:Canet:2003yu,
GasenzerRef:Latorre:2004pk,
GasenzerRef:Kehrein2004a,
GasenzerRef:Mitra2006a,
GasenzerRef:Zanella:2006am,
GasenzerRef:Canet:2006xu,
GasenzerRef:Gezzi2007a,
GasenzerRef:Jakobs2007a,
GasenzerRef:Korb2007a,
GasenzerRef:Matarrese:2007wc,
GasenzerRef:Karrasch2008a,
GasenzerRef:Jakobs2009a,
GasenzerRef:Schoeller2009a}.
To obtain a practically solvable set of dynamic equations, one derives the flow equation for the proper $n$-point Green function $\Gamma_{\tau}^{(n)}$ by taking the $n$th field derivative of \GasenzerCmdEq{flowGamma}.  
This scheme yields a set of coupled integro-differential equations for the $\Gamma_{\tau}^{(n)}$, and eventually for the connected $n$-point functions $G^{(n)}$, including the Schwinger-Dyson equation for the two-point Green function $G$.
Its power lies in both a physically motivated evaluation of the effective action and the possibility to derive non-perturbative equations of motion in a compact form \cite{GasenzerRef:Gasenzer:2008zz,GasenzerRef:Gasenzer:2010rq}.

\section{The 2PI effective action beyond mean-field order}
\label{GasenzerRef:sec:2PITrunc}
In order to evaluate the dynamic equation \GasenzerCmdeq{EOMG}, details about the self-energy $\Sigma$ are required, and these are, in all interesting cases, only available to a certain approximation. 
The natural expansion of $\Gamma_2$ is in terms of 2PI closed loop diagrams involving only bare vertices and full propagators $G$.
This expansion can be truncated at any order, e.g., of powers of the bare coupling $g$ or of the number of loops.

\subsection{Mean-field and quantum Boltzmann truncations}
\label{GasenzerRef:sec:Interrelations}
Mean-field approximations that lead to the Gross-Pitaevskii and the Hartree-Fock-Bogoliubov (HFB) equations emerge as leading-order truncations of the 2PI effective action ($\Gamma_2=0$, and $\Gamma_2=$ double-bubble ${\cal O}(g)\text{ diagram}$, respectively, see \GasenzerCmdFig{Gamma23loop}).
Higher truncations take into account non-mean-field effects of the collisional interactions.
The closest connection to well-known formulations which account for scattering is obtained within the leading-order truncation in powers of the bare coupling $g$, taking into account the ``Basketball'' diagram in $\Gamma_{2}$.
At this level of approximation the resulting dynamic equations can be reduced, after introducing quasi-particles at the mean-field level, and a Markov approximation within the scattering integral, to the well-known quantum Boltzmann equations for the single-particle densities, see, e.g., Ref.~\cite{GasenzerRef:Branschadel:2008sk}.
The Basketball diagram pictorially illustrates that it describes two-to-two scattering with momentum conservation, and that it is of second order in the coupling.

\subsection{Conservation laws and long-time stability}
\label{GasenzerRef:sec:Applicability}
The most prominent advantage of the 2PI approach over, e.g., equations derived from the 1PI effective action or the BBGKY hierarchy of coupled equations for cumulants, is that, whatever truncation is chosen, the resulting dynamic equations respect the conservation of energy and particle number.
This property is due to the self-consistent determination of $G$, and $\phi$ in the case of bosons.
As a result, provided an implementation with sufficient numerical precision, the equations do not lead to a secular evolution or the emergence of negative occupation numbers.
This is particularly important for approximations which include the effect of scattering and allow to describe long-time evolution including equilibration and thermalization \cite{GasenzerRef:Berges:2004yj}.

\subsection{Non-perturbative expansions beyond mean-field order}
\label{GasenzerRef:sec:NLO1N}
While $\mathcal{O}(g^n)$ truncations of $\Gamma_{2}$ render the approach useful for perturbative approximations beyond mean-field-order, resummations of infinite classes of diagrams are possible which extend the applicability beyond the coupling expansion.
In the following, we will review a few applications of a resummation procedure corresponding to the next-to-leading order (NLO) truncation of an expansion of $\Gamma_{2}$ in inverse powers of the number of internal field degrees of freedom $\mathcal{N}$ \cite{GasenzerRef:Berges:2001fi,GasenzerRef:Aarts:2002dj}, see \GasenzerCmdFig{Gamma23loop} and Refs.~\cite{GasenzerRef:Berges:2002cz,GasenzerRef:Mihaila:2000sr,GasenzerRef:Cooper:2002qd,GasenzerRef:Arrizabalaga:2004iw,GasenzerRef:Berges:2002wr,GasenzerRef:Berges:2004ce,GasenzerRef:Berges:2004pu,GasenzerRef:Berges:2004yj,GasenzerRef:Aarts:2006pa,GasenzerRef:Gasenzer:2005ze,GasenzerRef:Rey2005a,GasenzerRef:Temme2006a,GasenzerRef:Berges:2007ym,GasenzerRef:Branschadel:2008sk,GasenzerRef:Berges:2008wm,GasenzerRef:Scheppach:2009wu,GasenzerRef:Sexty:2010ra} for non-equilibrium applications.
This procedure takes into account classical statistical fluctuations to infinite order in the coupling  and includes quantum fluctuations to leading order in the quantum contribution to the interaction vertex \cite{GasenzerRef:Berges:2007ym}.
Hence, in this truncation, the dynamic equations are particularly useful when studying the dynamics of systems in which quantum statistical fluctuations are suppressed.
This is, e.g., generically the case in Bose gases at low energies in which the bulk of particles occupies a few excited modes near zero energy.
In the context of Kadanoff-Baym equations the NLO $1/\mathcal{N}$ expansion is known as GW-approximation \cite{GasenzerRef:Hedin1965a,GasenzerRef:Aryasetiawan1998a} and has been the subject of non-equilibrium studies recently \cite{GasenzerRef:vanLeeuwen2006a,GasenzerRef:Myohanen2008a,GasenzerRef:Myohanen2009a}.

\section{Applications}
\label{GasenzerRef:sec:Apps}
We present three examples where the dynamic equations derived from a non-perturbatively approximated 2PI effective action provide insight into the many-body time evolution beyond the mean-field as well as leading-order quantum-Boltzmann equations.

\subsection{1D lattice gas: 2PI versus exact dynamics}
\label{GasenzerRef:sec:Josephson}
In the first example, we consider a coupled few-mode Bose system as, e.g., a  one-dimensional (1D) lattice gas in the tight-binding approximation. 
Its full quantum dynamics can be computed exactly or simulated in the quasiclassical regime, and we compare such results with predictions obtained within the 2PI approach.
The system is defined by the Bose-Hubbard Lagrangian 
$
  {\cal L}(n,t)
=
  \left[\frac{i}{2}\Psi_n^*(t)\partial_{t}\Psi_n(t)
     +J\Psi_n^*(t)\Psi_{n+1}(t)+\text{c.c.}\right]
     -\epsilon_n\Psi_n^*(t)\Psi_n(t)
     -\frac{U}{2}[\Psi_n^*(t)\Psi_n(t)]^2,
$
$n=1,...,N_{s}$ being the lattice site index. 
For $N_{s}=2$, the system of $N$ particles is equivalent to a Josephson junction with coupling energy $E_\mathrm{J}=2JN$  and charging energy $E_\mathrm{c}=2U$, cf., e.g., Refs.~\cite{GasenzerRef:Paraoanu2001a,GasenzerRef:Pitaevskii2001a}.
%
%
\begin{figure}[tb]
\begin{center}
\includegraphics[width=0.62\textwidth]{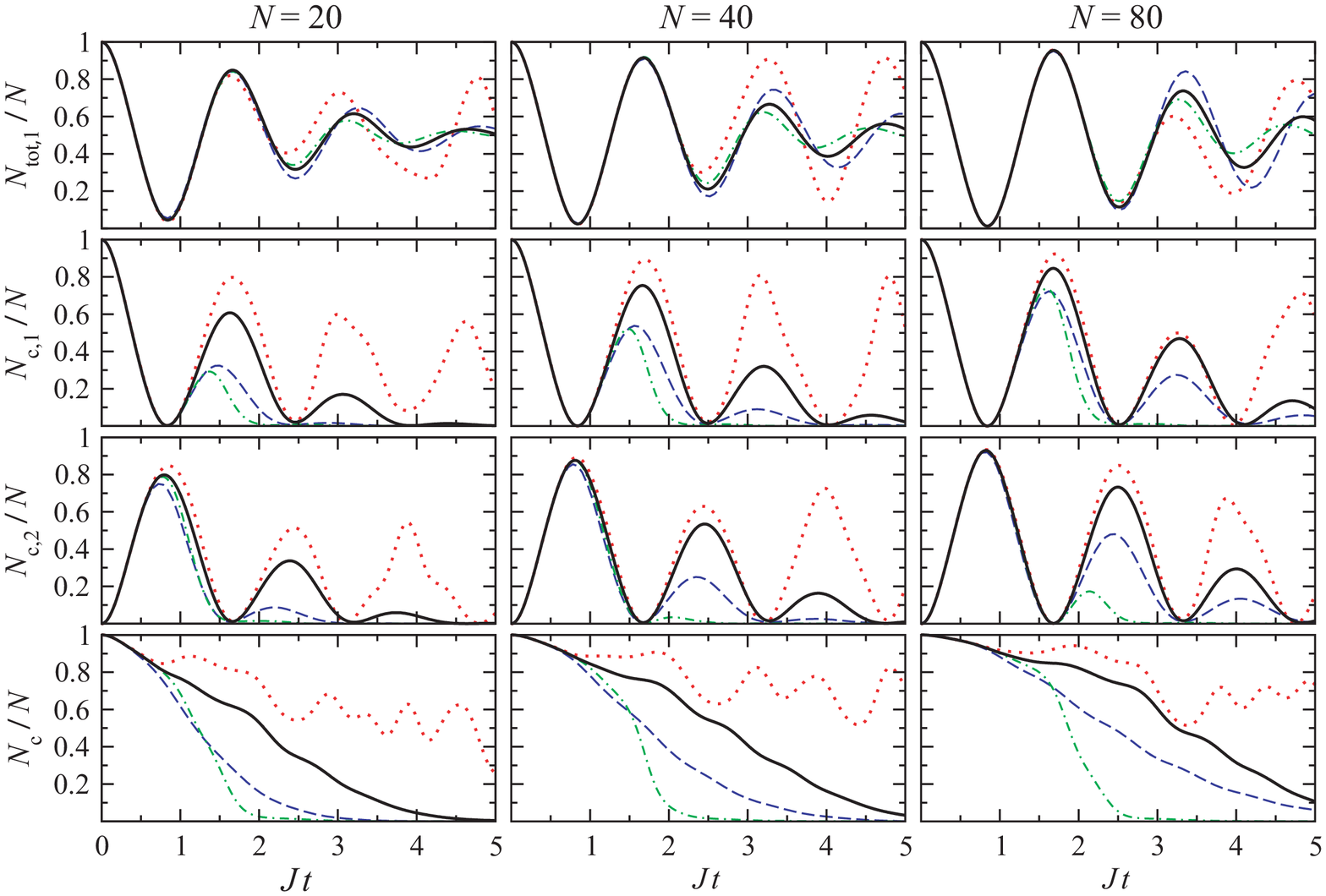}
\includegraphics[width=0.34\textwidth]{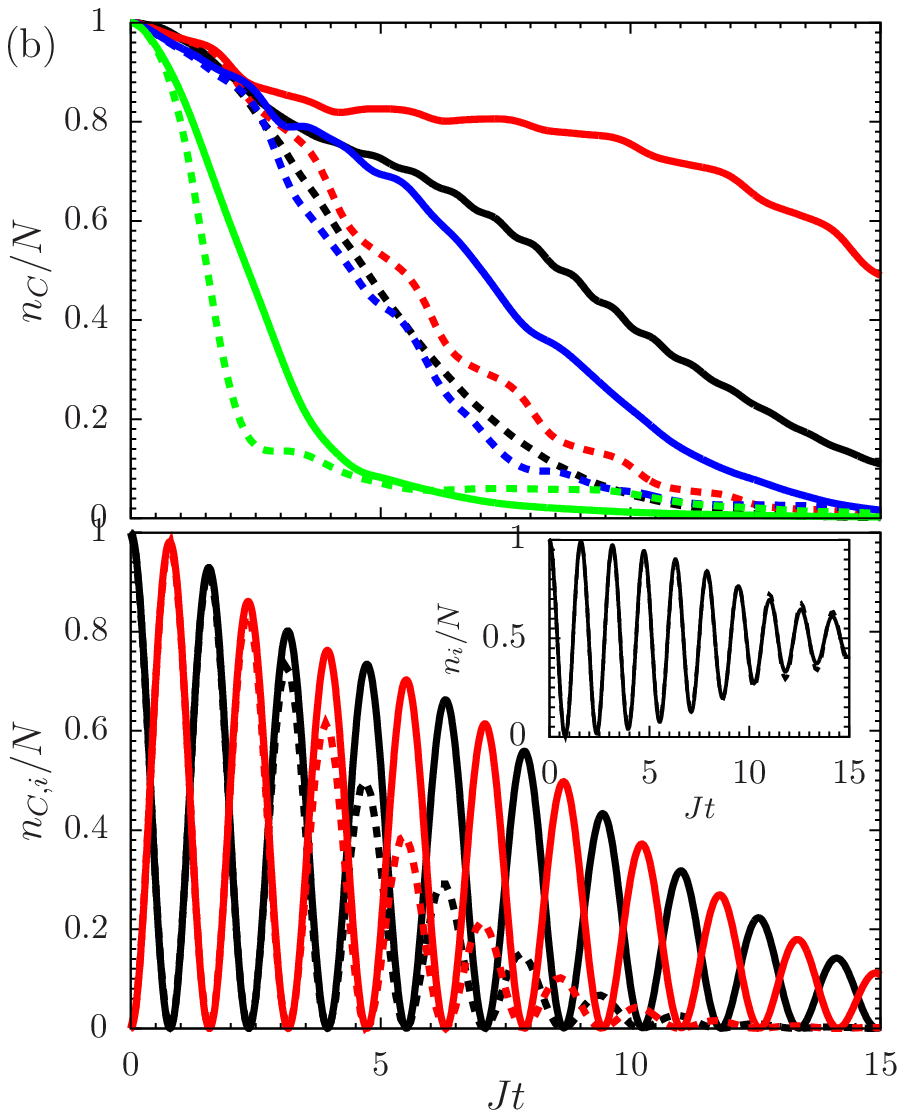}
\end{center}
\vspace*{-3ex}
\caption{
(a) Time evolution of the Josephson contact between $N_s=2$ bosonic modes.
The three columns show the same quantities for a different total number of atoms $N=20$, $40$, and $80$, respectively.
$NU/J\equiv4$ throughout.
Initially, all atoms are in a Bose-Einstein condensate in mode $1$.
First line: Total number of atoms $N_\mathrm{tot,1}$ in mode $1$.
Due to number conservation, $N_\mathrm{tot,2}=N-N_\mathrm{tot,1}$.
Second and third lines: Condensate fractions $N_{\mathrm{c},n}/N$ at sites $n=1,2$.
Fourth line: $N_\mathrm{c}/N=\sum_nN_{\mathrm{c},n}/N$.
$1/\cal N$ expansion to NLO: Thick solid line.
Blue dashed lines: Exact calculation.
Green dash-dotted curves: $1/\cal N$ expansion in 2nd-order coupling approximation.
Red dotted curves: Hartree-Fock-Bogoliubov approximation.
(b) Upper:
Time evolution similar to (a) for different numbers of sites $N_s=2$ (black), $3$ (red), $5$ (blue) and $10$ (green).
The other parameters are chosen as $NU/N_sJ=5$ and $N/N_s=10$.
Solid line: $1/\cal N$ expansion to NLO.
Dashed line: quasi-exact semi-classical simulations.
Lower:
Evolution of the condensate fraction and of the total population (inset) per site for $N_s=2$.
}
\label{GasenzerRef:fig:dynamics2}
\end{figure}
In order to probe the accuracy of the NLO $1/\cal N$ approximation, we chose $NU/J$ to be larger than $1$ and compare our results with the results of an exact numerical calculation.
In \GasenzerCmdFig{dynamics2}a, the 2PI evolution in NLO $1/\cal N$ approximation is compared to the HFB and second-order coupling approximations, see Ref.~\cite{GasenzerRef:Temme2006a} for further details.
The time evolution shows different characteristic periods.
At early times, the condensate Rabi oscillates coherently between the lattice wells with frequency $\Omega=2E_\mathrm{J}/N=4J$.
Only a small number of atoms is scattered from the condensate fraction into excited modes.
Lateron, atoms are exchanged between the condensate and the non-condensate modes of the gas.
These processes lead to a rapid destruction of the condensate fraction and to damping of the Rabi oscillations.

In \GasenzerCmdFig{dynamics2}b, we compare the evolution of the total condensate fraction according to the 2PI NLO $1/{\cal N}$ equations with the quasi-exact semi-classical simulations (``Truncated Wigner Approximation''), for $N_s$ up to 10. These results illustrate that the field-theoretic non-perturbative resummation is capable of qualitatively describing the evolution. 
Note that, in contrast to this, coupling expansions generically break down at large times. 
Quantitative agreement converges at most slowly with increasing number of sites.
We furthermore find that the total number of particles at a particular site is described much better, by the NLO equations, than the condensate fraction alone, which is strongly overestimated, see \GasenzerCmdFig{dynamics2}b for an example. This indicates that, in the NLO $1/{\cal N}$ approximation, the phase spreading in the different modes is captured only to a limited extent.

\subsection{Non-thermal equilibration of a 1D Fermi gas}
\label{GasenzerRef:sec:Fermi}
As a second example, we study the long-time evolution and equilibration of a 1D Fermi gas \cite{GasenzerRef:Kronenwett:2010dx,GasenzerRef:Kronenwett:2010ic}, containing two spin components $\alpha\in\{\uparrow,\downarrow\}$ that mutually interact through local repulsive $s$-wave collisions described by the Hamiltonian
$H=  \int \mathrm{dx}[\Psi^{\dagger}_{\alpha}(x)({-\partial_{\mathrm{x}}^{2}}/{2m})\Psi_{\alpha}(x)
  +({g_{\alpha\beta}}/{2})
    \Psi^{\dagger}_{\alpha}(x)\Psi^{\dagger}_{\beta}(x)
    \Psi_{\beta}(x)\Psi_{\alpha}(x)],
$
with $g_{\alpha\beta}=(1-\delta_{\alpha\beta})\,4\pi a_{\mathrm{1D}}/m$, $a_{\mathrm{1D}}$ being the 1D scattering length.
This model is considered integrable in the sense that it has as many conserved quantities as there are degrees of freedom \cite{GasenzerRef:Yang1967a}.
As a consequence, the gas is expected not to be described by a grand-canonical density matrix at large times \cite{GasenzerRef:Rigol2007a,GasenzerRef:Manmana2007a,GasenzerRef:Gangardt2008a}.

An extensive discussion has recently focused on the question under which circumstances a generalized Gibbs ensemble (GGE) can describe the long-time properties of, in particular, integrable systems \cite{GasenzerRef:Rigol2007a,GasenzerRef:Calabrese2007a,GasenzerRef:Kollath2007a,GasenzerRef:Eckstein2008a, 
GasenzerRef:Moeckel2008a,GasenzerRef:Meden2010a}.
From the experimental point of view, it appears difficult to decide on which general grounds one may be in the position to falsify the equilibration to a particular final density matrix.
We will show that, instead of restricting a measurement to a single point in time, it is useful to look for a possible deviation from the fluctuation-dissipation theorem. 
The results indicate that a degenerate 1D Fermi gas defined by the Hamiltonian $H$ approaches a non-thermal state at large times.

%
\begin{figure*}[tb]
{ \centering
  \includegraphics[width=.98\textwidth]{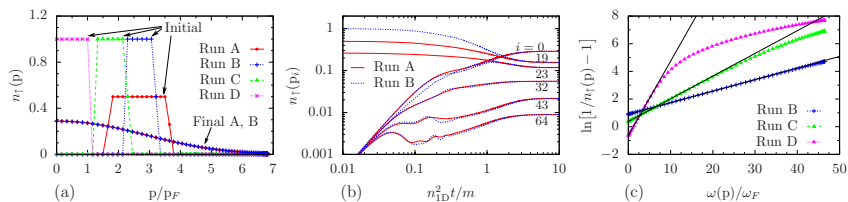}
}
  \caption{ \label{GasenzerRef:fig:RunAAndB} 
    Equilibration of a one-dimensional Fermi gas starting from different initial momentum distributions.
    (a)
    Initial ($t = 0$) and final ($t=10\,mn_{\text{1D}}^{-2}$) distributions $n_\uparrow(t,|\mathrm{p}|)= n_\downarrow(t,|\mathrm{p}|)$ in runs A and B have the same total particle number and energy.
    In runs C and D, the particle number is the same while the energies are lower than in B.
    (b)
    Occupation numbers $n_\uparrow(t,|\mathrm{p}|)$ as a function of time $t$ for momentum modes $\mathrm{p}_{i}=\sin[i\pi/N_{s}]/a_{s}$,
    $N_s = 128$.
    (c)
    Inverse-slope functions $\sigma=\ln[1/n_\uparrow(p)-1]$ for runs B, C, and D at late times.
    The black dashed lines are Fermi-Dirac fits to the lowest 14 momentum modes from which the temperatures and chemical potentials are extracted.
  }
\end{figure*}
\begin{figure}[tb]
  \includegraphics[width=.98\textwidth]{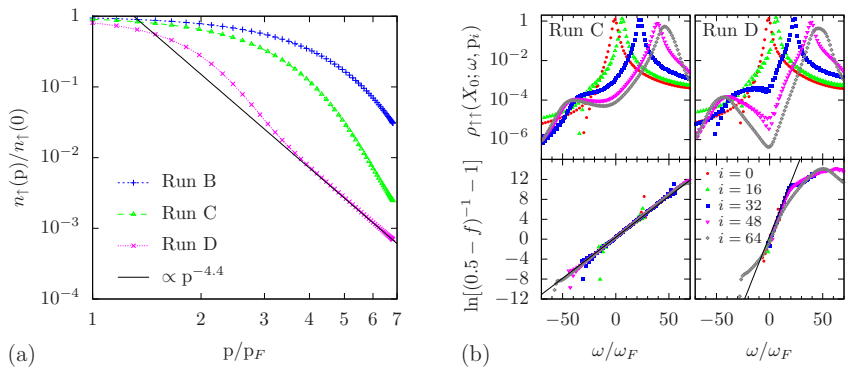}
  \caption{  \label{GasenzerRef:fig:FDT} 
  (a)
  Late-time momentum distributions equivalent to those shown in \GasenzerCmdFig{RunAAndB}c.
  (b)
  Upper row: Spectral functions as a function of frequency at late time $X_{0}=18.9\,n_{\text{1D}}^{-2}m$ in runs C and D for five of the momentum modes $\mathrm{p}_{i}$, see legend in lower right panel.
  Lower row: Inverse-slope function of fractions $f$ of the statistical correlation function $F$ divided by the spectral function $\rho$ at $X_{0}=18.9\,n_{\text{1D}}^{-2}m$ for the same five momentum modes, see main text.
  Black lines indicate Fermi-Dirac distributions with the same $\beta$ and $\mu$ as in \GasenzerCmdFig{RunAAndB}c.
  In run D, the system does not thermalize.
}
\end{figure}
%
%
In \GasenzerCmdFig{RunAAndB}b, the time-evolution of six single-particle momentum modes $n_\uparrow(t,|\mathrm{p}_{i}|)= n_\downarrow(t,|\mathrm{p}_{i}|)$ are shown, for an unpolarized gas following an interaction quench at $t_{0}$ from $\gamma=mg/n_{\mathrm{1D}}=0$ to $\gamma=4$,
for different initial momentum distributions $n_\uparrow(|\mathrm{p}|)=n_\uparrow(t_{0},|\mathrm{p}|)$ (Runs A and B in \GasenzerCmdFig{RunAAndB}a).
The equilibration process shown in \GasenzerCmdFig{RunAAndB}b is characterized by a short-time scale of dephasing depending on the width of the initial momentum distribution and a long-time scale determined by scattering.
The different initial distributions of runs A and B have the same particle number and energy such that both runs approach the same final state.
The initial momentum distributions in runs C and D further shown in \GasenzerCmdFig{RunAAndB}a contain the same particle number as runs A and B but have lower energy.
We find that also runs C and D reach stationary momentum distributions which,  however, are no longer given by a Fermi-Dirac function over the entire range of momenta.
As seen in \GasenzerCmdFig{RunAAndB}c, the lower momentum modes appear thermalized while the higher momenta in runs C and D remain overpopulated as compared to the exponential fall-off of the Fermi-Dirac distribution.

\GasenzerCmdFig{FDT}a shows the $n_\uparrow(|\mathbf{p}|)$
for runs C and D on a double-logarithmic scale.
The high-momentum tails are characterized by a power-law $n(\mathrm{p})\propto \mathrm{p}^{-\kappa}$ with $\kappa\simeq 4.4$.
This exponent does not change when we consider truncations of the full 2PI loop expansion that contain all diagrams up to order $g^2$ or $g^3$.
This non-Fermi-Dirac momentum distribution in itself is not a proof for non-thermal equilibration.
However, the final state is found to also violate the fluctuation dissipation theorem which points to a non-(grand-)canonical character. 
Given a grand-canonical many-body state $\rho=Z^{-1}\exp[\beta(H-\mu N)]$ where $N$ is the total-particle-number operator, the fluctuation-dissipation theorem states that the statistical correlation function $F_{\alpha\alpha}(X_{0};\omega,\mathrm{p})=\int \mathrm{d}s \exp(i\omega s)F_{\alpha\alpha}(X_{0}+s/2, X_{0}-s/2;$ $\mathrm{p})$, where 
$F_{\alpha\alpha}(t,t';\mathrm{p})=\langle[\Psi^\dagger_{\alpha}(t,\mathrm{p}),\Psi_{\alpha}(t',\mathrm{p})]\rangle/2$, and the spectral function  $\rho_{\alpha\alpha}(t,t';\mathrm{p})=i\langle\{\Psi^\dagger_{\alpha}(t,\mathrm{p}),\Psi_{\alpha}(t',\mathrm{p})\}\rangle$ are related by
\begin{equation}
  F_{\alpha\alpha}(X_{0};\omega,\mathrm{p})
   = -i[ 1/2-n_\mathrm{FD}(\omega-\mu) ]
     \rho_{\alpha\alpha}(X_{0};\omega,\mathrm{p}).
\end{equation}
In \GasenzerCmdFig{FDT}, we depict, at the late time $X_{0}=18.9\,n_{\text{1D}}^{-2}m$, the fraction $f=i F_{\uparrow\uparrow}(X_{0};\omega,\mathrm{p})/\rho_{\uparrow\uparrow}(X_{0};\omega,\mathrm{p})$ as a function of the frequency $\omega$ for five different momentum modes $\mathrm{p}_{i}$.
The lower left (right) panel shows the inverse-slope function $\ln[(1/2-f)^{-1}-1]$ of $f$ for run C (D).
$f$ is shown in a region including the spectral peaks where the argument of the logarithm is positive.
In run C, over the region of relevant $\omega$, this function is a straight line corresponding to a Fermi-Dirac function $f$.
Hence, the system is thermalized over the depicted range of energies, despite the power-law tail in run C found for $\omega_{\mathrm{p}}>30\,\omega_{F}$, see \GasenzerCmdFig{RunAAndB}c.
This can be understood by considering the spectral function in \GasenzerCmdFig{FDT}b (upper row) on a logarithmic scale.
A second peak at negative frequencies, not present in an ideal gas, picks up extra contributions from the Fermi sea.
It is this second peak which causes the power-law overpopulation at high momenta, in the same way as of the BCS zero-temperature depletion of the Fermi sea in a weakly interacting system.
Hence, in run C, the system thermalizes to a state within a grand-canonical ensemble, with the eigenmodes of the strongly interacting system being superpositions of particles and holes.
Run D, however, performed at even lower energy, shows that the system does in general not thermalize to a grand-canonical ensemble.
As shown in the lower right panel of \GasenzerCmdFig{FDT}b, the result of this run violates the fluctuation-dissipation theorem.
Although the momentum overpopulation is again largely produced by the contributions from the Fermi sea, also the fraction $f$ shows a power-law tail $\sim \mathrm{p}^{-9}$.
In conclusion, the 2PI equations describe non-thermal equilibration of a 1D Fermi gas with positive contact interactions, at sufficiently low total  energies, to a state violating the fluctuation-dissipation theorem for a grand-canonical ensemble.

\subsection{Non-thermal fixed points and quantum turbulence}
\label{sec:QT}
In the last example, we discuss non-equilibrium fixed points of a Bose gas.
The dynamic equations derived from the non-perturbatively approximated 2PI effective action can be analyzed with respect to stationary scaling solutions.
The idea is to ask for a solution of the dynamic equation \GasenzerCmdeq{EOMG} that is translation-invariant in time and space, and is a scaling function in the following way: The frequency-momentum dependent statistical function (see previous section) obeys $F(s^{z}q_{0},s\mathbf{q})=s^{-2-\kappa}F(q_{0},\mathbf{q})$, while the spectral function satisfies $\rho(s^{z}q_{0},s\mathbf{q})=s^{-2}\rho(q_{0},\mathbf{q})$, the latter in accordance with the commutation relations.
The scaling law for $F$ implies the single-particle spectrum $n(\mathbf{k})$ to scale as
\begin{equation}
n(s \mathbf{k}) = s^{-\zeta}n(\mathbf{k}),
\end{equation}
where $s$ is some positive real number and $\zeta=\kappa+2-z$.
Within kinetic theory, where the quantum Boltzmann equation takes the role of the equation of motion for $n(\mathbf{k},t)$, scaling laws can be derived by use of Zacharov integral transformations, `little miracles of wave-turbulence theory' \cite{GasenzerRef:Zakharov1992a}. 
These transformations allow to rewrite the complicated scattering integral in such a way that scaling exponents of solutions other than the Rayleigh-Jeans distribution $n\sim T/k^2$ can be read off.

To determine the positive exponent $\zeta$ in the infrared (IR) regime, where $n(\mathbf{k})\sim| \mathbf{k}|^{-\zeta}$ is large, an approach beyond kinetic theory is required.
Recall that the `Basketball' diagram in the 2nd line of \GasenzerCmdFig{Gamma23loop} (left panel) gives the quantum Boltzmann equation.
This diagram is proportional to a lower power of the occupation number $n$ than all higher-order ones.
Hence, the approximation becomes unreliable for $n\gg1$.
In contrast, the NLO $1/\mathcal{N}$ approximation or $s$-channel resummation involves diagrams to all orders in the coupling and in $n$.
While, in the regime of large wave numbers, the approach goes over into the kinetic description of `weak wave turbulence', an effectively renormalized many-body $T$-matrix modifies the scaling law in the IR.
Details of this and of the derivation of the scaling exponent $\zeta$ are given in Ref.~\cite{GasenzerRef:Scheppach:2009wu}.

\begin{figure}[!t]
 \includegraphics[width=0.48\textwidth]{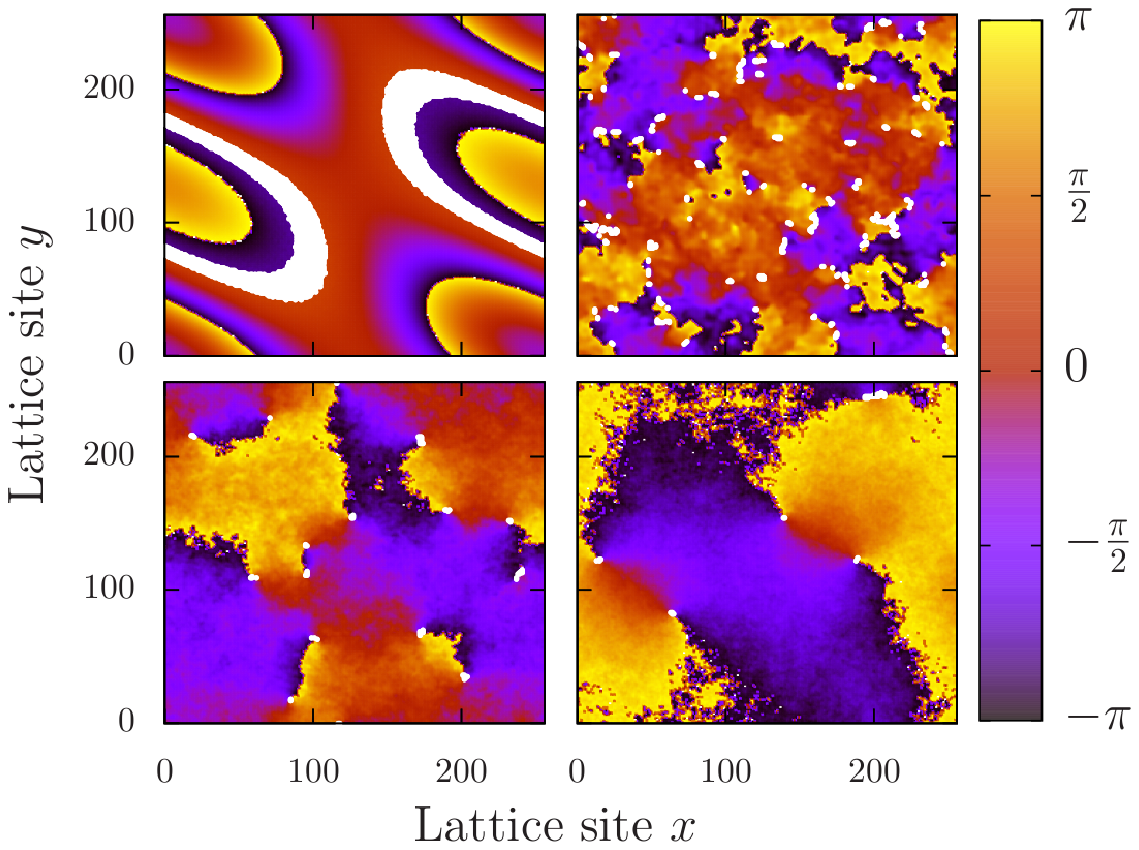}\hspace{0.02\textwidth}
 \includegraphics[width=0.48\textwidth]{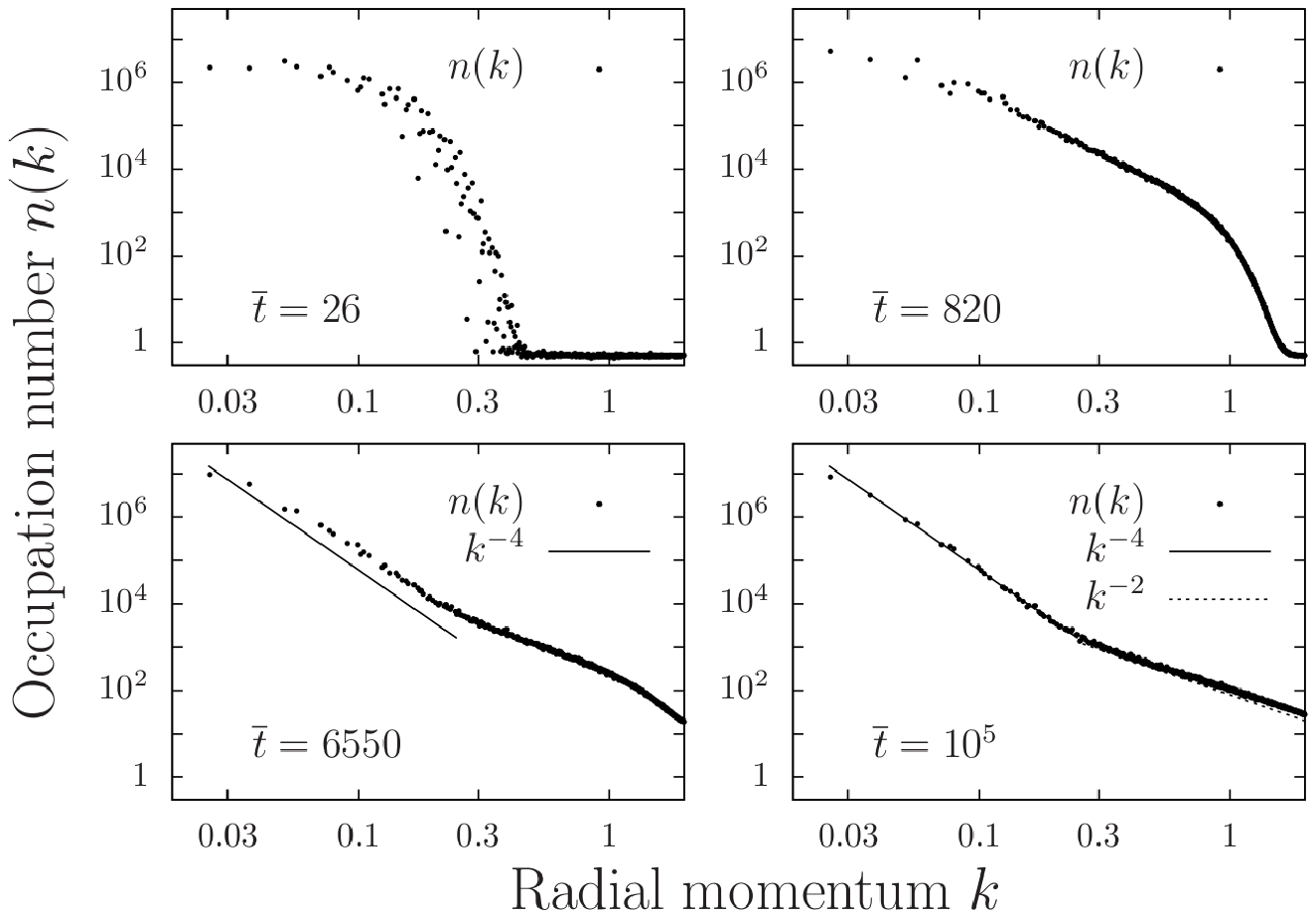}
 \caption{Left panel: Phase angle (color scale) at four times during a single run of the simulations in $2$ dimensions. 
The white spots mark vortex cores where the density falls below 5\% of the mean density. 
Shown times are (in lattice units): 
1. $\overline{t}=26$ (top left): Ordered phase shortly after initial preparation. 
2. $\overline{t}=820$ (top right): After creation of vortex-antivortex pairs. 
3. $\overline{t}=6550$ (bottom left): During critical slowing down of the vortex-antivortex annihilation. 
4.  $\overline{t}=10^5$ (bottom right): Low-density vortex-antivortex pairs before final thermalization.
Right panel:
Mode occupation numbers as functions of the radial momentum $k$, for the four different times of the run in $2$ dimensions shown on the left.
Note the double-logarithmic scale.
An initial broadening of the momentum distribution is followed by the early development of a scaling $n(k)\sim k^{-4}$ and the later emergence of the bimodal scaling with $n(k)\sim k^{-2}$ at larger wave numbers.
Scaling laws are indicated by thin lines.
\label{GasenzerRef:fig:QTurbulence}
}
\end{figure}
The renormalization of the coupling is, in the NLO $1/\mathcal{N}$-approximation, a consequence of the resummation of an infinite number of Feynman diagrams contributing to the 2PI effective action in terms of a geometric series \cite{GasenzerRef:Berges:2008wm,GasenzerRef:Berges:2008sr,GasenzerRef:Scheppach:2009wu}.
Physically, the renormalized $T$-matrix implies a reduction of the effective interaction strength in the IR regime of strongly occupied bosonic field modes \cite{GasenzerRef:Berges:2008wm}. 
As a consequence of the reduced interactions, single-particle occupation numbers rise towards smaller wave numbers in a steeper way than in the (ultraviolet) kinetic regime.
The following IR scaling exponents were predicted in Ref.~\cite{GasenzerRef:Scheppach:2009wu}:
\begin{align}
  \zeta = \zeta^\mathrm{IR}_{\mathrm{Q}}
  &= d+2 ,
  \label{GasenzerRef:eq:kappaIRQ}
  \\
  \zeta = \zeta^\mathrm{IR}_{\mathrm{P}}
  &= d+2+z 
  \label{GasenzerRef:eq:kappaIRP}
\end{align}
where $d$ is the number of spatial dimensions,  
$\omega \sim k^z$, and Q (P) indicates that the quasiparticle number (energy) flows at constant rate in the cascade. 

Recently, these scaling exponents where shown, by means of simulations of the Gross-Pitaevskii equation, to be closely related to the phenomenon of quantum turbulence \cite{GasenzerRef:Nowak:2010tm}.
Given a generic out-of-equilibrium initial condition for an ultracold quantum gas, topological features such as quantized vortices or vortex lines may appear after a short evolution time.
They are usually produced in great numbers, in 2D together with antivortices, to comply with the overall angular momentum of the system, in 3D in the form of closed lines.
While diminishing again in number due to pair annihilation or reconnection, clear power laws appear in the occupation number distribution, confirming the predictions obtained from the non-perturbative 2PI approach, see \GasenzerCmdFig{QTurbulence}.
See Ref.~\cite {GasenzerRef:Berges:2010ez} for a large-$\mathcal{N}$ relativistic study.

\section{Summary}
\label{GasenzerRef:sec:Summary}
We have briefly reviewed functional-integral approaches to non-equilibrium quantum many-body dynamics, including the 2PI-effective action and the functional flow-equation method.
These techniques provide a powerful and economic access to dynamics of strongly correlated systems, both analytically and numerically.
Approximations far beyond the mean-field level conserve crucial quantities like energy and particle number and allow to describe long-time dynamics and equilibration.
With a few example applications we illustrated the potential of the approach, demonstrating the description of equilibration both to thermal and non-thermal states, the evolution of uniform as well as inhomogeneous systems, of Bosons and Fermions.
Also specific properties of non-equilibrium stationary states like critical scaling can be described, and the comparison with alternative methods proves that the methods are useful for studying non-trivial topological solutions.

\section*{Acknowledgments}

The authors acknowledge support by the Deutsche Forschungsgemeinschaft, by the Alliance Programme of the Helmholtz Association (HA216/EMMI), by the Excellence Programme FRONTIER of the University of Heidelberg, by the German Academic Exchange Service (DAAD), by the Heidelberg Graduate School for Fundamental Physics, and by the Landesgraduiertenf\"orderung Baden-W\"urttemberg.
T. G. thanks EPSRC for support concerning FINESS 2009.



\begin{thebibliography}{47}
\providecommand{\natexlab}[1]{#1}
\providecommand{\url}[1]{\texttt{#1}}
\expandafter\ifx\csname urlstyle\endcsname\relax
  \providecommand{\doi}[1]{doi: #1}\else
  \providecommand{\doi}{doi: \begingroup \urlstyle{rm}\Url}\fi

\bibitem{GasenzerRef:Luttinger1960a}
J.~M. Luttinger and J.~C. Ward, Ground-state energy of a many-fermion system.
  {II}, \emph{Phys. Rev.} {\bf 118}, \penalty0 1417 (1960).

\bibitem{GasenzerRef:Baym1962a}
G.~Baym, Self-consistent approximations in many-body systems, \emph{Phys. Rev.}
  {\bf 127}, \penalty0 1391 (1962).

\bibitem{GasenzerRef:Cornwall1974a}
J.~M. Cornwall, R.~Jackiw, and E.~Tomboulis, Effective action for composite
  operators, \emph{Phys. Rev. D}. {\bf 10}, \penalty0 2428 (1974).

\bibitem{GasenzerRef:Vasiliev1998a}
A.~N. Vasiliev, \emph{Functional Methods in Quantum Field Theory and
  Statistical Physics}. (Gordon and Breach, Amsterdam, 1998).

\bibitem{GasenzerRef:Kleinert1982a}
H.~Kleinert, Higher effective actions for {B}ose systems, \emph{Fortschr.
  Phys.} {\bf 30}, \penalty0 187  (1982).

\bibitem{GasenzerRef:Berges:2001fi}
J.~Berges, Controlled nonperturbative dynamics of quantum fields out of
  equilibrium, \emph{Nucl. Phys.} {\bf A699}, \penalty0 847  (2002).

\bibitem{GasenzerRef:Aarts:2002dj}
G.~Aarts, D.~Ahrensmeier, R.~Baier, J.~Berges, and J.~Serreau,
  Far-from-equilibrium dynamics with broken symmetries from the 2{PI}-1/{N}
  expansion, \emph{Phys. Rev. D}. {\bf 66}, \penalty0 045008  (2002).

\bibitem{GasenzerRef:Berges:2002cz}
J.~Berges and J.~Serreau, Parametric resonance in quantum field theory,
  \emph{Phys. Rev. Lett.} {\bf 91}, \penalty0 111601  (2003).

\bibitem{GasenzerRef:Mihaila:2000sr}
B.~Mihaila, F.~Cooper, and J.~F. Dawson, {Resumming the large-N approximation
  for time evolving quantum systems}, \emph{Phys. Rev. D}. {\bf 63}, \penalty0
  096003  (2001).

\bibitem{GasenzerRef:Cooper:2002qd}
F.~Cooper, J.~F. Dawson, and B.~Mihaila, Quantum dynamics of phase transitions
  in broken symmetry {$\lambda\phi^4$} field theory, \emph{Phys. Rev. D}. {\bf
  67}, \penalty0 056003  (2003).

\bibitem{GasenzerRef:Arrizabalaga:2004iw}
A.~Arrizabalaga, J.~Smit, and A.~Tranberg, Tachyonic preheating using {2PI} -
  {1/N} dynamics and the classical approximation, \emph{JHEP}. {\bf 10},
  \penalty0 017  (2004).

\bibitem{GasenzerRef:Berges:2002wr}
J.~Berges, S.~Borsanyi, and J.~Serreau, Thermalization of fermionic quantum
  fields, \emph{Nucl. Phys.} {\bf B660}, \penalty0 51  (2003).

\bibitem{GasenzerRef:Berges:2004ce}
J.~Berges, S.~Borsanyi, and C.~Wetterich, Prethermalization, \emph{Phys. Rev.
  Lett.} {\bf 93}, \penalty0 142002  (2004).

\bibitem{GasenzerRef:Berges:2004yj}
J.~Berges, Introduction to nonequilibrium quantum field theory, \emph{AIP Conf.
  Proc.} {\bf 739}, \penalty0 3,  \emph{arXiv:hep-ph/0409233} (2005).

\bibitem{GasenzerRef:Aarts:2006pa}
G.~Aarts and J.~M. Martinez~Resco, {Transport coefficients from the 2PI
  effective action: Weak coupling and large N analysis}, \emph{J. Phys. Conf.
  Ser.} {\bf 35}, \penalty0 414  (2006).

\bibitem{GasenzerRef:Berges:2008wm}
J.~Berges, A.~Rothkopf, and J.~Schmidt, {Non-thermal fixed points: effective
  weak-coupling for strongly correlated systems far from equilibrium},
  \emph{Phys. Rev. Lett.} {\bf 101}, \penalty0 041603  (2008).

\bibitem{GasenzerRef:Berges:2004pu}
J.~Berges, n-PI effective action techniques for gauge theories, \emph{Phys.
  Rev. D}. {\bf 70}, \penalty0 105010 (2004).

\bibitem{GasenzerRef:Rey2004a}
A.~Rey, B.~Hu, E.~Calzetta, A.~Roura, and C.~Clark, Nonequilibrium dynamics of
  optical-lattice-loaded {B}ose-{E}instein-condensate atoms: Beyond the
  {H}artree-{F}ock-{B}ogoliubov approximation, \emph{Phys. Rev. A}. {\bf 69},
  \penalty0 033610 (2004).

\bibitem{GasenzerRef:Baier:2004hm}
R.~Baier and T.~Stockamp, Kinetic equations for {B}ose-{E}instein condensates from
  the {2PI} effective action.  \emph{arXiv:hep-ph/0412310} (2004).

\bibitem{GasenzerRef:Gasenzer:2005ze}
T.~Gasenzer, J.~Berges, M.~G. Schmidt, and M.~Seco, Non-perturbative dynamical
  many-body theory of a {B}ose-{E}instein condensate, \emph{Phys. Rev. A}. {\bf
  72}, \penalty0 063604  (2005).

\bibitem{GasenzerRef:Rey2005a}
A.~M. Rey, B.~L. Hu, E.~Calzetta, and C.~W. Clark, Quantum kinetic theory of a
  {B}ose-{E}instein gas confined in a lattice, \emph{Phys. Rev. A}. {\bf 72},
  \penalty0 023604 (2005).

\bibitem{GasenzerRef:Temme2006a}
K.~Temme and T.~Gasenzer, Non-equilibrium dynamics of condensates in a lattice
  from the {2PI} effective action in {$1/{\cal N}$} expansion, \emph{Phys. Rev.
  A}. {\bf 74}, \penalty0 053603  (2006).

\bibitem{GasenzerRef:Berges:2007ym}
J.~Berges and T.~Gasenzer, Quantum versus classical statistical dynamics of an
  ultracold {B}ose gas, \emph{Phys. Rev. A}. {\bf 76}, \penalty0 033604 (2007).

\bibitem{GasenzerRef:Branschadel:2008sk}
A.~Bransch{\"{a}}del and T.~Gasenzer, {2PI} nonequilibrium versus transport
  equations for an ultracold {B}ose gas, \emph{J. Phys. B}. {\bf 41}, \penalty0
  135302  (2008).

\bibitem{GasenzerRef:Scheppach:2009wu}
C.~Scheppach, J.~Berges, and T.~Gasenzer, Matter-wave turbulence: Beyond
  kinetic scaling, \emph{Phys. Rev. A}. {\bf 81}\penalty0 (3), \penalty0 033611
  (2010).

\bibitem{GasenzerRef:Sexty:2010ra}
D.~Sexty, T.~Gasenzer, and J.~Pawlowski, {Real-time effective-action approach
  to the {A}nderson quantum dot}, \emph{arXiv:1012.4293 [cond-mat.mes-hall]}  (2010).

\bibitem{GasenzerRef:Wetterich:1992yh}
C.~Wetterich, Exact evolution equation for the effective potential, \emph{Phys.
  Lett.} {\bf B301}, \penalty0 90  (1993).

\bibitem{GasenzerRef:Bagnuls:2000ae}
C.~Bagnuls and C.~Bervillier, Exact renormalization group equations: An
  introductory review, \emph{Phys. Rept.} {\bf 348}, \penalty0 91  (2001).

\bibitem{GasenzerRef:Berges:2000ew}
J.~Berges, N.~Tetradis, and C.~Wetterich, Non-perturbative renormalization flow
  in quantum field theory and statistical physics, \emph{Phys. Rept.} {\bf
  363}, \penalty0 223  (2002).

\bibitem{GasenzerRef:Polonyi:2001se}
J.~Polonyi, Lectures on the functional renormalization group method,
  \emph{Central Eur. J. Phys.} {\bf 1}, \penalty0 1  (2003).

\bibitem{GasenzerRef:Salmhofer:2001tr}
M.~Salmhofer and C.~Honerkamp, Fermionic renormalization group flows: Technique
  and theory, \emph{Prog. Theor. Phys.} {\bf 105}, \penalty0 1  (2001).

\bibitem{GasenzerRef:Delamotte:2003dw}
B.~Delamotte, D.~Mouhanna, and M.~Tissier, Nonperturbative renormalization
  group approach to frustrated magnets, \emph{Phys. Rev. B} {\bf 69}, \penalty0
  134413  (2004).

\bibitem{GasenzerRef:Salmhofer:2006pn}
M.~Salmhofer, {Dynamical Adjustment of Propagators in Renormalization Group
  Flows}, \emph{Annalen Phys.} {\bf 16}, \penalty0 171  (2007).

\bibitem{GasenzerRef:Gies:2006wv}
H.~Gies, {Introduction to the functional {RG} and applications to gauge
  theories}, \emph{arXiv:hep-ph/0611146}  (2006).

\bibitem{GasenzerRef:Igarashi:2009tj}
Y.~Igarashi, K.~Itoh, and H.~Sonoda, {Realization of symmetry in the {ERG}
  approach to quantum field theory}, \emph{Prog. Theor. Phys. Suppl.} {\bf
  181}, \penalty0 1  (2010).

\bibitem{GasenzerRef:Rosten:2010vm}
O.~J. Rosten, {Fundamentals of the Exact Renormalization Group},
  \emph{arXiv:1003.1366 [hep-th]}  (2010).

\bibitem{GasenzerRef:Litim:1998nf}
D.~F. Litim and J.~M. Pawlowski, On gauge invariant {W}ilsonian flows. \emph{arXiv:hep-th/9901063} (1999).

\bibitem{GasenzerRef:Litim:2006ag}
D.~F. Litim and J.~M. Pawlowski, Non-perturbative thermal flows and
  resummations, \emph{JHEP}. {\bf 11}, \penalty0 026  (2006).

\bibitem{GasenzerRef:Pawlowski:2005xe}
J.~M. Pawlowski, Aspects of the functional renormalisation group, \emph{Annals
  Phys.} {\bf 322}, \penalty0 2831  (2007).

\bibitem{GasenzerRef:Canet:2003yu}
L.~Canet, B.~Delamotte, O.~Deloubriere, and N.~Wschebor, Non perturbative
  renormalization group study of reaction- diffusion processes and directed
  percolation, \emph{Phys. Rev. Lett.} {\bf 92}, \penalty0 195703 (2004).

\bibitem{GasenzerRef:Latorre:2004pk}
J.~I. Latorre, C.~A. Lutken, E.~Rico, and G.~Vidal, Fine-grained entanglement
  loss along renormalization group flows, \emph{Phys. Rev. A}. {\bf 71},
  \penalty0 034301  (2005).

\bibitem{GasenzerRef:Kehrein2004a}
S.~Kehrein, Scaling and decoherence in the nonequilibrium {K}ondo model,
  \emph{Phys. Rev. Lett.} {\bf 95}, \penalty0 056602  (2005).

\bibitem{GasenzerRef:Mitra2006a}
A.~Mitra, S.~Takei, Y.~B. Kim, and A.~J. Millis, Nonequilibrium quantum
  criticality in open electronic systems, \emph{Phys. Rev. Lett.} {\bf 97},
  \penalty0 236808  (2006).

\bibitem{GasenzerRef:Zanella:2006am}
J.~Zanella and E.~Calzetta, Renormalization group study of damping in
  nonequilibrium field theory, \emph{hep-th/0611222}  (2006).

\bibitem{GasenzerRef:Canet:2006xu}
L.~Canet and H.~Chate, {Non-perturbative Approach to Critical Dynamics},
  \emph{J. Phys. A}. {\bf 40}, \penalty0 1937  (2007).

\bibitem{GasenzerRef:Gezzi2007a}
R.~Gezzi, T.~Pruschke, and V.~Meden, Functional renormalization group for
  nonequilibrium quantum many-body problems, \emph{Phys. Rev. B}. {\bf
  75}\penalty0 045324 (2007).

\bibitem{GasenzerRef:Jakobs2007a}
S.~G. Jakobs, V.~Meden, and H.~Schoeller, Nonequilibrium functional
  renormalization group for interacting quantum systems, \emph{Phys. Rev.
  Lett.} {\bf 99}, \penalty0 150603 (2007).

\bibitem{GasenzerRef:Korb2007a}
T.~Korb, F.~Reininghaus, H.~Schoeller, and J.~K{\"{o}}nig, Real-time
  renormalization group and cutoff scales in nonequilibrium, \emph{Phys. Rev.
  B}. {\bf 76}\penalty0 165316  (2007).

\bibitem{GasenzerRef:Matarrese:2007wc}
S.~Matarrese and M.~Pietroni, Resumming cosmic perturbations, \emph{JCAP}. {\bf
  0706}, \penalty0 026  (2007).

\bibitem{GasenzerRef:Karrasch2008a}
C.~Karrasch, R.~Hedden, R.~Peters, T.~Pruschke, K.~Sch{\"{o}}nhammer, and
  V.~Meden, A finite-frequency functional {RG} approach to the single impurity
  Anderson model, \emph{J. Phys.: Cond. Mat.} {\bf 20}, \penalty0
  345205  (2008).

\bibitem{GasenzerRef:Jakobs2009a}
S.~G. Jakobs, M.~Pletyukhov, and H.~Schoeller.
\newblock Nonequilibrium functional RG with frequency dependent vertex function
  -- a study of the single impurity {A}nderson model, 
\emph{J. Phys. A: Math. Theor.} {\bf 43}, \penalty0
  103001   (2010).

\bibitem{GasenzerRef:Schoeller2009a}
H.~Schoeller, {A perturbative nonequilibrium renormalization group method for
  dissipative quantum mechanics. Real-time {RG} in frequency space
  ({RTRG-FS})}, \emph{Eur. Phys. J. ST}. {\bf 168}, \penalty0 179  (2009).

\bibitem{GasenzerRef:Gasenzer:2008zz}
T.~Gasenzer and J.~M. Pawlowski, Functional renormalisation group approach to
  far-from-equilibrium quantum field dynamics, \emph{Phys. Lett.} {\bf B670}, 135  (2008).

\bibitem{GasenzerRef:Gasenzer:2010rq}
T.~Gasenzer, S.~Kessler, and J.~M. Pawlowski, {Far-from-equilibrium quantum
  many-body dynamics}, \emph{Eur. Phys. J. C}. {\bf 70}, \penalty0 423
  (2010).

\bibitem{GasenzerRef:Hedin1965a}
L.~Hedin, New method for calculating the one-particle {G}reen's function with
  application to the electron-gas problem, \emph{Phys. Rev.} {\bf 139}\penalty0
  (3A), \penalty0 A796 (1965).

\bibitem{GasenzerRef:Aryasetiawan1998a}
F.~Aryasetiawan and O.~Gunnarsson, The {GW} method, \emph{Rep. Prog. Phys.}
  {\bf 61}, \penalty0 237  (1998).

\bibitem{GasenzerRef:vanLeeuwen2006a}
R.~van Leeuwen, N.~E. Dahlen, and A.~Stan, Total energies from variational
  functionals of the {G}reen function and the renormalized four-point vertex,
  \emph{Phys. Rev. B}. {\bf 74}\penalty0 (19), \penalty0 195105 (2006).

\bibitem{GasenzerRef:Myohanen2008a}
P.~My\"oh\"anen, A.~Stan, G.~Stefanucci, and R.~van Leeuwen, A many-body
  approach to quantum transport dynamics: Initial correlations and memory
  effects, \emph{Eur. Phys. Lett.} {\bf 84}, \penalty0 67001  (2008).

\bibitem{GasenzerRef:Myohanen2009a}
P.~My\"oh\"anen, A.~Stan, G.~Stefanucci, and R.~van Leeuwen, {K}adanoff-{B}aym
  approach to quantum transport through interacting nanoscale systems: From the
  transient to the steady-state regime, \emph{Phys. Rev. B}. {\bf 80}\penalty0
  (11), \penalty0 115107 (2009).

\bibitem{GasenzerRef:Paraoanu2001a}
G.-S. Paraoanu, S.~Kohler, F.~Sols, and A.~J. Leggett, The {J}osephson plasmon
  as a {B}ogoliubov quasiparticle, \emph{J. Phys. B: At. Mol. Opt. Phys.} {\bf
  34}, \penalty0 4689  (2001).

\bibitem{GasenzerRef:Pitaevskii2001a}
L.~Pitaevskii and S.~Stringari, Thermal vs quantum decoherence in double well
  trapped {B}ose-{E}instein condensates, \emph{Phys. Rev. Lett.} {\bf
  87}\penalty0 (18), \penalty0 180402 (2001).

\bibitem{GasenzerRef:Kronenwett:2010dx}
M.~Kronenwett and T.~Gasenzer, Nonthermal equilibration of a one-dimensional
  {F}ermi gas, \emph{arXiv:1006.3330 [cond-mat.quant-gas]}  (2010).

\bibitem{GasenzerRef:Kronenwett:2010ic}
M.~Kronenwett and T.~Gasenzer, Far-from-equilibrium dynamics of an ultracold
  fermi gas, \emph{arXiv:1012.3874 [cond-mat.quant-gas]}  (2010).

\bibitem{GasenzerRef:Yang1967a}
C.~N. Yang, Some exact results for the many-body problem in one dimension with
  repulsive delta-function interaction, \emph{Phys. Rev. Lett.} {\bf
  19}\penalty0 (23), \penalty0 1312 (1967).

\bibitem{GasenzerRef:Rigol2007a}
M.~Rigol, V.~Dunjko, V.~Yurovsky, and M.~Olshanii, Relaxation in a completely
  integrable many-body quantum system: An ab initio study of the dynamics of
  the highly excited states of 1d lattice hard-core bosons, \emph{Phys. Rev.
  Lett.} {\bf 98}, \penalty0 050405 (2007).

\bibitem{GasenzerRef:Manmana2007a}
S.~R. Manmana, S.~Wessel, R.~M. Noack, and A.~Muramatsu, Strongly correlated
  fermions after a quantum quench, \emph{Phys. Rev. Lett.} {\bf 98}, \penalty0
  210405 (2007).

\bibitem{GasenzerRef:Gangardt2008a}
D.~M. Gangardt and M.~Pustilnik, Correlations in an expanding gas of hard-core
  bosons, \emph{Phys. Rev. A}. {\bf 77}, \penalty0 041604(R)  (2008).

\bibitem{GasenzerRef:Calabrese2007a}
P.~Calabrese and J.~Cardy, Quantum quenches in extended systems, \emph{J. Stat.
  Mech.} {\bf P06008}  (2007).

\bibitem{GasenzerRef:Kollath2007a}
C.~Kollath, A.~M. L\"auchli, and E.~Altman, Quench dynamics and nonequilibrium
  phase diagram of the bose-hubbard model, \emph{Phys. Rev. Lett.} {\bf
  98}\penalty0 (18), \penalty0 180601 (2007).

\bibitem{GasenzerRef:Eckstein2008a}
M.~Eckstein and M.~Kollar, Nonthermal steady states after an interaction quench
  in the falicov-kimball model, \emph{Phys. Rev. Lett.} {\bf 100}\penalty0
  (12), \penalty0 120404 (2008).

\bibitem{GasenzerRef:Moeckel2008a}
M.~Moeckel and S.~Kehrein, Interaction quench in the hubbard model, \emph{Phys.
  Rev. Lett.} {\bf 100}\penalty0 (17), \penalty0 175702 (2008).

\bibitem{GasenzerRef:Meden2010a}
D.~M. Kennes and V.~Meden, Relaxation dynamics of an exactly solvable
  electron-phonon model, \emph{Phys. Rev. B}. {\bf 82}\penalty0 (8), \penalty0
  085109 (2010).

\bibitem{GasenzerRef:Zakharov1992a}
V.~E. Zakharov, V.~S. {L'vov}, and G.~Falkovich, \emph{Kolmogorov Spectra of
  Turbulence I: Wave Turbulence}. (Springer-Verlag, Berlin, 1992).

\bibitem{GasenzerRef:Berges:2008sr}
J.~Berges and G.~Hoffmeister, {Nonthermal fixed points and the functional
  renormalization group}, \emph{Nucl. Phys.} {\bf B813}, \penalty0 383
  (2009).

\bibitem{GasenzerRef:Nowak:2010tm}
B.~Nowak, D.~Sexty, and T.~Gasenzer, Quantum turbulence in an ultracold {B}ose
  gas, \emph{arXiv:1012.4437 [cond-mat.quant-gas]}  (2010).

\bibitem{GasenzerRef:Berges:2010ez}
J.~Berges and D.~Sexty, {Strong versus weak wave-turbulence in relativistic
  field theory}, \emph{arXiv:1012.5944 [hep-ph]}  (2010).

\end{thebibliography}
\end{document}